\documentclass[12pt]{article}
\usepackage{epsf}
\usepackage{graphicx}

\begin{document}
\title
{The graviton background: a new way to quantum gravity}
\author
{Michael A. Ivanov \\
Physics Dept.,\\
Belarus State University of Informatics and Radioelectronics, \\
6 P. Brovka Street,  BY 220027, Minsk, Republic of Belarus.\\
E-mail: ivanovma@gw.bsuir.unibel.by.}

\maketitle

\begin{abstract}Graviton pairing and destruction of these pairs
under collisions with bodies may lead to the Newtonian attraction.
It opens us a new way to a very-low-energy quantum gravity model.
In the model by the author, cosmological redshifts are caused by
interactions of photons with gravitons of the background.
Non-forehead collisions with gravitons lead to an additional
relaxation of any photonic flux. Total galaxy number
counts/redshift and galaxy number counts/magnitude relations are
computed and found to be in a good agreement with galaxy
observations.
\end{abstract}
\section[1]{Where does the land of quantum gravity lie?}
After creation of quantum mechanics, many attempts were made to
understand a quantum side of gravity. With time, a commonly
accepted now opinion has been summed that quantum gravity is a
high-energy phenomenon which should manifest itself only on a huge
scale of energies of the order of the Planck energy $\sim 10^{19}$
GeV. An important difference of existing models of quantum gravity
from other quantum models of elementary particle physics is their
primordially accepted {\it geometrical} base. Perhaps, many
difficulties are conditioned namely by this circumstance. Poorness
of the set of predicted or expected effects \cite{33} is casted to
an eye on the "output" of such the models, and I think that the
very eloquent assertion by S. Carlip may summarize the situation:
"... there is not yet any direct experimental evidence that
gravity is quantized" \cite{33}.
\par Models with large extra dimensions \cite{81,82,83} lead us to
TeV-scale gravity, and, maybe, they appropinquate us to a real
energy scale of quantum gravity. But the notion that all our world
is not any more than a slice of a real and non-available (as a
whole) for us world is not very reassuring. Extra dimensions are
{\it necessary} to describe, for example, the composite
fundamental fermions \cite{7} when they may by interpreted as
internal coordinates of the composite system, but their
assignement to any "empty" point of space seems to me to be an
excess.
\par But there exists another side of medal. Nesvizhevsky's team
reported about discovery of quantum states of ultra-cold neutrons
in the Earth's gravitational field \cite{4}. Observed energies of
levels (it means that and their differences too) in full agreement
with quantum-mechanical calculations turned out to be equal to
$\sim 10^{-12}$ eV. If transitions between these levels are
accompanied with irradiation of gravitons then energies of
irradiated gravitons should have the same order that is of 40
orders lesser than the Planck energy. Anderson's team reported
about the anomalous acceleration of NASA's probes Pioneer 10/11
\cite{1}; this effect is not embedded in a frame of the general
relativity, and its magnitude is somehow equal to $\sim Hc$, where
$H$ is the Hubble constant, $c$ is the light velocity. In the same
1998, two teams of astrophysicists reported about dimming remote
supernovae \cite{2,3}; the one would be explained on a basis of
the Doppler effect if at the present epoch the universe expands
with acceleration. This explanation needs an introduction of some
"dark energy" which is unknown from any laboratory experiment. But
the last effect as well as the Pioneer anomaly may be explained as
the additional manifestations of low-energy quantum gravity based
on the idea of an existence of the graviton background
\cite{5,500}. The main results of author's research in this
approach are considered here (for more details, see \cite{500}).

\section[2]{Do we live in the sea of gravitons?}
We should observe the following effects in the see of gravitons: a
redshift due to forehead collisions of photons with gravitons and
an additional relaxation of any light flux due to their
none-forehead collisions \cite{500}. The first effect leads to the
geometrical distance/redshift relation of the form: $r(z)= ln
(1+z)/a,$ where $a=H/c,$ $H$ is the Hubble constant. The both
effects give the luminosity distance $D_{L}(z):$ $D_{L}(z)=a^{-1}
\ln(1+z)\cdot (1+z)^{(1+b)/2}.$ This function fits supernova
observations very well for roughly $z < 0.5,$ that excludes a need
of any dark energy to explain supernovae dimming. Total galaxy
number counts $dN(z) \propto f_{2}(z)= {ln^{2}(1+z) \over
{z^{2}(1+z)}}$ (see \cite{90}). A graph of this function is shown
in Fig. 1; the typical error bar and data point are added here
from paper \cite{72} by Loh and Spillar. There is not a visible
contradiction with observations.
\begin{figure}[th]
\epsfxsize=12.98cm \centerline{\epsfbox{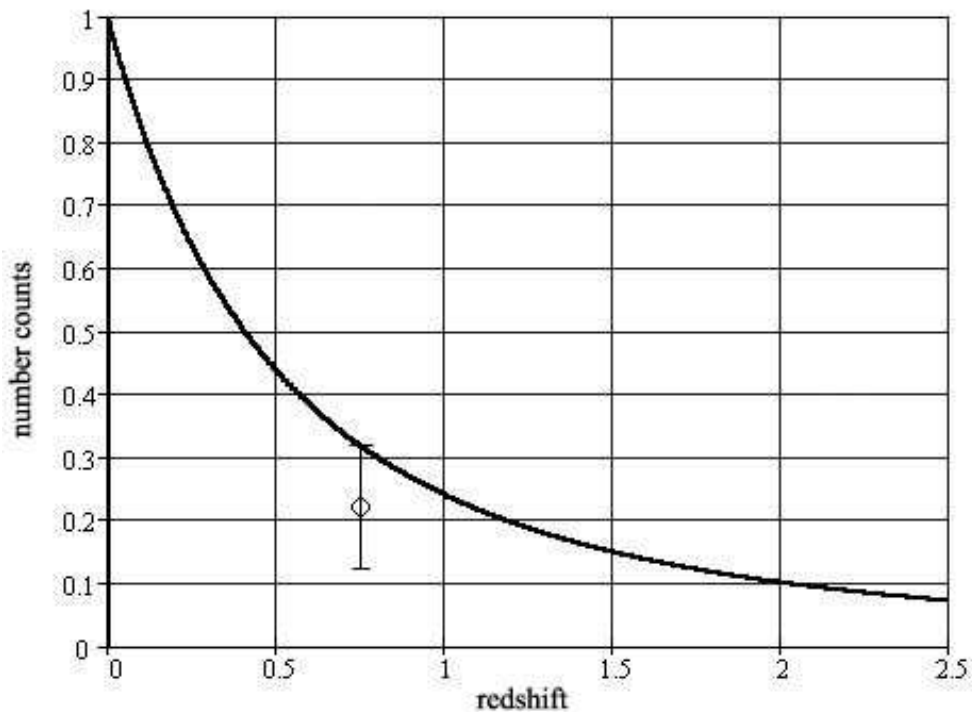}} \caption{Number
counts $f_{2}$ as a function of the redshift in this model. The
typical error bar and data point are taken from paper \cite{72} by
Loh and Spillar.}
\end{figure}
Usually, one ascribes the decrease of galaxy number counts per a
"unit volume" to an expansion of the Universe. But {\it there is
not any expansion in this model}.
\par
The galaxy number counts/magnitude relation $f_{3}(m),$ $m$ is a
magnitude, in this model (see \cite{90}), which takes into account
the Schechter luminosity function \cite{73}, gives us another
possibility to test a reality of the graviton background. To
compare this function with observations by Yasuda et al.
\cite{77}, we can choose the normalizing factor from the
condition: $f_{3}(16)=a(16), $ where $a(m)\equiv A_{\lambda}\cdot
10^{0.6(m-16)}$ is the function giving the best fit to
observations \cite{77}, $A_{\lambda}=const.$ In this case, we have
two free parameters of the Schechter distribution - $\alpha$ and
$L_{\ast}$ - to fit observations, and the latter one is connected
with a constant $A_{1}\simeq 5\cdot 10^{17}\cdot {L_{\odot} /
L_{\ast}}.$ The ratio ${{f_{3}(m) -a(m)}\over a(m)}$ is shown in
Fig.2 for different values of $A_{1}$ by $\alpha =-2.43.$
\begin{figure}[th]
\epsfxsize=12.98cm \centerline{\epsfbox{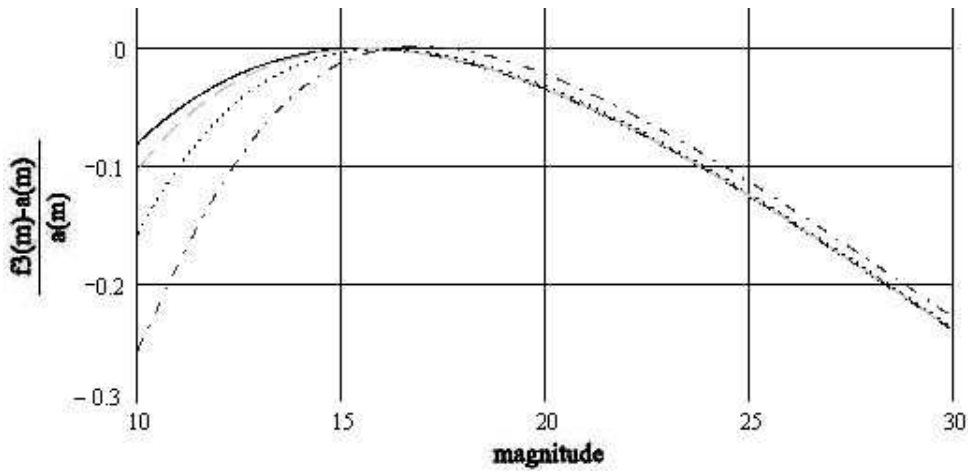}}
\caption{The relative difference $(f_{3}(m)-a(m))/a(m)$ as a
function of the magnitude $m$ for $\alpha=-2.43$ by
$10^{-2}<A_{1}<10^{2}$ (solid), $A_{1}=10^{4}$ (dash),
$A_{1}=10^{5}$ (dot), $A_{1}=10^{6}$ (dadot). }
\end{figure}
If we compare this figure with Figs. 6,10,12 from \cite{77}, we
see that the considered model provides a no-worse fit to
observations than the function $a(m)$ if the same K-corrections
are added (perhaps, even a better one if one takes into account
positions of observational points in these figures by $m<16$ and
$m>16$).
\par
In all of the three cases - supernova dimming, total galaxy number
counts/redshift and galaxy number counts/magnitude relations - we
can see a good enough accordance of the theory with observations.
\section[3]{The main manifestation of the graviton background:
the Newtonian attraction} The single graviton screening creates
{\it equal} attractive and repulsive forces for any pair of usual
bodies \cite{500}. The forces are very big, about three order
greater than the Newtonian force. The net force cannot be equal to
zero in a case of black hole, and we have immediately a very
important consequence: {\it the equivalence principle should be
violated if black holes exist.}
\par It was shown by the author that graviton pairing may ensure
the Newtonian attraction. In this model, the cross-section of
interaction depends bilinear on particle energies; then, after
destruction of graviton pairs in a collision with a body, single
scattered gravitons give a repulsive force which is twice smaller
than an attractive one due to a pressure of pairs. It turns out
that one should consider a space distribution of gravitons and
their pairs in a realization of flat wave to compute the Newton
constant. Today this model is a semi-classical one, but it is an
underlaying one relative to the Newtonian limit and the general
relativity (the Newton constant is {\it computable} here). It is
interesting and absolutely unexpected that the model needs an
"atomic structure" of matter to be {\it workable}. The following
condition of big distances is necessary to use a geometrical
language in gravity: $\sigma (E,<\epsilon_{2}>) \ll 4 \pi r^{2},$
where $\sigma (E,<\epsilon_{2}>)$ is the cross-section of
interaction of a particle with an energy $E$ and a graviton pair
with an average energy $<\epsilon_{2}>$, $r$ is a distance between
them. For a proton, this condition leads to the restriction: $r
\gg 10^{-11} \ m$. This limit length is many orders greater than
the Planck scale of $\sim 10^{-33} \ m$. By distances smaller than
$10^{-11} \ m,$ the equivalence principle should be broken. This
prediction may mean that any consideration of gravity in a
geometrical language is nonsense in this forbidden scale.
\section[4]{Conclusion}
In the considered model, we start from a micro level but we see
immediately such the very important and long time observed
cosmological effect as redshifts of remote objects. A deceleration
of massive bodies (of the order of $Hc$), when they move relative
to the graviton background, is an analog of this effect. A similar
deceleration has been observed only for the Pioneer 10/11
spacecrafts \cite{1}, and the problem is open what is happen for
big bodies. During almost a century (from 1922, when redshifts
were first observed), nobody thought that the cosmological
redshift may be considered as an effect of low-energy quantum
gravity. I would like to hope that in the future it will be
recognized as the one in a shorter time.

\end{document}